\documentclass[11pt, a4paper]{article}
\usepackage[T1]{fontenc}
\usepackage{inputenc}
\usepackage{geometry}
\usepackage{color}
\usepackage{amsmath}
\usepackage{amssymb}
\usepackage{cancel}
\usepackage{esint}
\usepackage{breakurl}

\usepackage{slashed}
\usepackage{cite}

\usepackage{jheppub}

\usepackage[x11names]{xcolor}
\usepackage{fancyhdr}
\usepackage{graphicx}
\usepackage{pgfplots}
\usepackage{tikz}

\makeatother

\global\long\def\e{\varepsilon}
 \global\long\def\del{\partial}
 \global\long\def\d{\delta}
 \global\long\def\grad{\nabla}

 \global\long\def\path{\mathcal{D}}

 \global\long\def\susn{\mathcal{N}}

\title{Magnetoresistance in relativistic hydrodynamics without anomalies}

\author[a]{Andrew Baumgartner,}
\author[a]{Andreas Karch}
\author[b]{and Andrew Lucas}

\affiliation[a]{Department of Physics, University of Washington, Seattle,
WA, 98195-1560, USA}
\affiliation[b]{Department of Physics, Stanford University, Stanford, CA,
94305-4060, USA}

\emailAdd{baum4157@uw.edu}
\emailAdd{akarch@uw.edu}
\emailAdd{ajlucas@stanford.edu}

\abstract{
We present expressions for the magnetoconductivity and the magnetoresistance
 of a strongly interacting metal in $3+1$ dimensions, derivable from relativistic hydrodynamics.   Such an approach is suitable for ultraclean metals with emergent Lorentz invariance.   When this relativistic fluid contains chiral anomalies, it is known to exhibit longitudinal negative magnetoresistance.  We show that similar effects can arise in non-anomalous relativistic fluids due to the distinctive gradient expansion.   In contrast with a Galilean-invariant fluid, the resistivity tensor of a dirty relativistic fluid exhibits similar angular dependence to negative magnetoresistance, even when the constitutive relations and momentum relaxation rate are isotropic.   We further account for the effect of magnetic field-dependent corrections to the gradient expansion and the effects of long-wavelength impurities on magnetoresistance.    We note that the holographic D3/D7 system exhibits negative magnetoresistance.
}

\begin{document}
\maketitle

\section{Introduction}

The behavior of metals in the presence of external electromagnetic
fields is of fundamental importance to our understanding of materials
and their transport properties. One such property is the magnetoresistance
(MR), which characterizes magnetic field dependence of the conductivity
tensor. For most metals, the longitudinal conductivity is a monotonically
decreasing function of $B$, or the resistance is a monotonically
increasing function of $B$ \cite{pippard_magnetoresistance_1989}.
However, the recent discovery of Weyl semimetals--materials in which
conduction bands intersect at distinct points in the Brillioun zone--shows
that this is not always the case \cite{li_chiral_2016,li_giant_2015,li_negative_2015,kim_dirac_2013,hirschberger_chiral_2016,huang_observation_2015,xiong_evidence_2015,zhang_signatures_2016}.

In Weyl semi-metals, quasi-particles with momentum values near the
intersection points (Weyl points) are described by the massless Weyl
Hamiltonian
\begin{equation}
H=\pm v\vec{\sigma}\cdot\left(-\mathrm{i}\grad-e\vec{A}\right)\label{eq:diracHAM}
\end{equation}
where $v$ is the velocity, $\vec{\sigma}$ is the pseudospin operator,
and $\pm$ correspond to the chirality of the quasi-particle. It is
a well known result in quantum field theory that in chiral theories
such as (\ref{eq:diracHAM}), the chiral (axial) current of these
fermions is not conserved. Such an anomaly, commonly referred to as
the Adler-Bell-Jackiw (ABJ) or chiral anomaly, is due to the breaking
of chiral symmetry by the path integral measure \cite{peskin_introduction_1996}.
Although this effect is manifestly quantum mechanical, it has important
consequences for classical transport. In particular, chiral fermions
can be spontaneously created if parallel electric and magnetic fields
are applied to the sample:
\begin{equation}
\del_{\mu}j_{\mathrm{A}}^{\mu}=-\frac{e^{2}}{8\pi^{2}}F_{\mu\nu}(\star F)^{\mu\nu}=\frac{e^{2}}{2\pi^{2}}\vec{E}\cdot\vec{B}\label{eq:divax}
\end{equation}
The subscript A on this current emphasizes that it is the axial current.

However, it is a theorem \cite{Nielsen:1983rb} that on a lattice, one must have
multiple Weyl points, such that the net chirality of the system vanishes.
Because these Weyl points are located a finite distance away from
each other in the Brillouin zone, there is always some non-vanishing
scattering rate for quasiparticles between the two Weyl points. Hence,
one must schematically modify (\ref{eq:divax}) to
\begin{equation}
\del_{\mu}j_{\mathrm{A}}^{\mu}=\frac{e^{2}}{2\pi^{2}}\vec{E}\cdot\vec{B}-\frac{\chi_{\mathrm{A}}\delta\mu_{\mathrm{A}}}{\tau},\label{eq:divax}
\end{equation}
where $\tau$ is the scattering time for quasiparticles to scatter
from the neighborhood of one Weyl point to another, and $\chi_{\mathrm{A}}\delta\mu_{\mathrm{A}}=\delta\rho_{\mathrm{A}}$
is the deviation in the axial charge density from equilibrium. Applying
an infinitesimal electric field $E_{i}$, and using Ohm's law
\begin{equation}
J_{i}=\sigma_{ij}E_{j}\label{eq:ohm}
\end{equation}
to extract the electrical conductivity, one finds\footnote{The anomaly itself drives a charge separation which yields an axial
charge growing linearly with time and so without any mechanisms to
release axial charge would result in run-away behavior. In the presence
of an relaxation mechanism for axial charge with time scale $\tau$,
one finds a finite build-up of axial charge $\delta\mu_{\mathrm{A}}$,
which in turn drives an electric current proportional to $B$ via
the chiral magnetic effect.}
\begin{equation}
\sigma_{ij}=\sigma_{0}\delta_{ij}+\frac{\tau e^{2}}{4\pi\chi_{\mathrm{A}}}B_{i}B_{j}.\label{eq:nmrAnom}
\end{equation}
In the limit $\tau\rightarrow\infty$, we find a parametrically large
contribution to the conductivity. The specific angular dependence
of this enhancement (only in the longitudinal direction oriented along
the magnetic field) is a key prediction.  It can be found within a kinetic description at weak coupling \cite{son_chiral_2013}, as well as a hydrodynamic \cite{lucas_hydrodynamic_2016, Roychowdhury:2015jha} or holographic \cite{jimenez-alba_anomalous_2015}  description at strong coupling.  
Evidence for such an angular structure was found in the recent
experiments \cite{li_chiral_2016,li_giant_2015,li_negative_2015,kim_dirac_2013,hirschberger_chiral_2016,huang_observation_2015,xiong_evidence_2015,zhang_signatures_2016}.

Already at weak coupling it has been demonstrated that NMR is possible
in non-anomalous systems \cite{goswami_axial_2015}. In the present
paper, we describe the hydrodynamic gradient expansion in background magnetic fields and derive a hydrodynamic equation for $\sigma_{ij}$, analogous to (\ref{eq:nmrAnom}), without any assumption of chirality.  We are inspired in part by
the holographic D3/D7 system, whose field theory dual is that of $\susn=2$
super Yang-Mills (SYM) hypermultiplets progagating through an $\susn=4$
SYM plasma \cite{ammon_holographic_2009}:  we will show that this system exhibits NMR.   We will also see that it is possible to obtain positive magnetoresistance within hydrodynamics, and will present two different `microscopic' mechanisms for this.

In fact, an order one contribution to the
current of the form $(\vec{E}\cdot\vec{B})B_{i}$ is generic.   Using symmetries
alone, and neglecting the breaking of rotational invariance by the microscopic crystal lattice, we anticipate the following expression for $\sigma_{ij}$:
\begin{equation}
\sigma_{ij}=\sigma_{0}\delta_{ij}+\sigma_{1}\epsilon_{ijk}B_{k}+\sigma_{2}B_{i}B_{j}.\label{eq:sigmasec1}
\end{equation}
The inverse of $\sigma_{ij}$,  the resistivity tensor $\rho_{ij}$, will also have similar structure: \begin{equation}
\rho_{ij}\equiv\alpha\delta_{ij}+\beta\epsilon_{ijk}B_{k}+\gamma\frac{B_{i}B_{j}}{B^{2}}.\label{eq:resist_tensor}
\end{equation}

In Section \ref{sec2}, we will show how (\ref{eq:resist_tensor}) generically
appears in relativistic hydrodynamics, with all $\alpha, \beta, \gamma \ne 0$, and relate these coefficients to the hydrodynamic dissipative coefficients.   An important difference between Galilean-invariant fluids and more general fluids (including relativistic fluids) is the fact that the charge current is not proportional to the momentum density in the latter case.   Hence, we will find that $\gamma \ne 0$, in contrast to the Galilean invariant case.   In fact, up to a brief discussion of thermal transport, our discussion of hydrodynamic charge transport is also valid for any non-Galilean invariant system.   

Nowhere do we assume the existence of any (emergent) axial anomaly.     Unlike in (\ref{eq:nmrAnom}), there is no
reason (a priori) to expect $\sigma_{2}$ to be parametrically large.   Still, we note that typical anomalous NMR observed
in experiment is not an order-of-magnitude enhancement.



Hydrodynamic transport in background fields has been applied successfully
to describe strongly correlated materials starting with the work of
\cite{hartnoll_theory_2007} on 2+1 dimensional physics. More recently,
this (relativistic) hydrodynamic approach to transport has also been
applied successfully to understand experimental transport data from
clean samples of graphene \cite{Crossno1058, Lucas:2015sya}. These studies take as input the
hydrodynamic transport coefficients, and then use the structure of
the hydrodynamic equations to give expressions for the
dependence of the transport properties on external fields and particle number
density. Starting from relativistic hydrodynamics instead of Galilean
hydrodynamics, one finds a number of distinct predictions in 2+1 dimensions
such as $B$-dependent lifetimes for cyclotron modes (a violation
of Kohn's theorem) \cite{hartnoll_theory_2007}.

One important aspect of this procedure is that one has to be careful
to work in a consistent expansion scheme. Hydrodynamics itself is
a gradient expansion, where $\partial_{\mu}$ is treated as a small
parameter. In the usual gradient expansion, since $\vec{B}=\nabla\times\vec{A}$,
the results of \cite{hartnoll_theory_2007} are, strictly speaking, only valid to linear
orders in $B$.
In this limit, one must treat multiple small parameters in the theory
as ``equally small" and only then perform the perturbative
expansion. If one wants to, for example, see the motion of hydrodynamic
poles (such as the cyclotron resonance) in the conductivity as a function
of $B$, one needs to include all higher transport coefficients involving
arbitrary powers of $B$. Since in our work all the interesting physics
appears at quadratic order in $B$, we will develop an expansion scheme
in which $B$ is considered to be zeroth order in derivatives  \cite{kovtun_thermodynamics_2016}.


The article is organized as followed: In Section \ref{sec2} we present
our hydrodynamic calculation, including the full conductivity and
resistivity tensors. We then discuss some interesting limits and their
physical interpretations. In Section \ref{sec3} we compare our results
to that of the D3/D7 system and a simple toy model made of ``electron-hole
plasma\textquotedbl{} in $3+1$ dimensions. The former is shown to
exhibit negative magnetoresistance, while the latter has positive
magnetoresistance.   We discuss a hydrodynamic model for the momentum relaxation time in Section \ref{sec4}, and show how the magnetic field generically leads to anisotropic momentum relaxation. We conclude in Section \ref{sec5}. 

As this paper was being finalized, \cite{Hernandez:2017mch} appeared, which contains some overlap with Section \ref{sec2}.

\section{Relativistic Hydrodynamics in a Magnetic Field}

\label{sec2}


\subsection{Weak Magnetic Fields}

Hydrodynamics is the low energy effective description of any interacting
quantum field theory, valid for fluctuations whose wavelength is much
larger than a `thermalization scale':  when quasiparticles are well-defined, this scale is simply the mean free path of the quasiparticles. When
we look on length scales long compared to the mean free path, the
system appears to be in local thermal equilibrium, thereby allowing
us to describe the global dynamics in terms of conserved quantities.
In this paper, these conserved quantities will  be charge,
energy and momentum. The dynamical equations in the presence of external
fields are \begin{subequations}\label{eq:momdisp}
\begin{gather}
\grad_{\mu}J^{\mu}=0\\
\grad_{\mu}T^{\mu\nu}=F^{\mu\nu}J_{\mu}+\frac{1}{\tau}\left(\d_{\:\nu}^{\mu}+u^{\mu}u_{\nu}\right)T^{\nu\lambda}u_{\lambda}
\end{gather}
\end{subequations} where the last term allows
for the dissipation of momentum due to impurities \cite{hartnoll_theory_2007};
it can be derived rigorously when the disorder strength is small from
multiple approaches \cite{rmp}.  The form of the $1/\tau$ term in (\ref{eq:momdisp}) is only sensible to linear order in the spatial components of $u^\mu$.  Following the conventions of \cite{hartnoll_theory_2007},
we can write the current and energy-momentum tensor as \begin{subequations}
\begin{gather}
J^{\mu}=\rho u^{\mu}+\nu^{\mu}+J_{\mathrm{mag}}^{\mu}\label{eq:curr}\\
T^{\mu\nu}=(\e+p)u^{\mu}u^{\nu}+pg^{\mu\nu}+\tau^{\mu\nu}+T_{\mathrm{mag}}^{\mu\nu}\label{eq:enTens}
\end{gather}
\end{subequations} where $\nu^{\mu}$ and $\tau^{\mu\nu}$ are dissipative
contributions which arise at first order in derivatives and a fluid
frame can be chosen in which they satisfy the following orthogonality
relations
\begin{equation}
u_{\mu}\nu^{\mu}=u_{\mu}\tau^{\mu\nu}=u_{\nu}\tau^{\mu\nu}=0.
\end{equation}
These terms can be found to first order in the derivative expansion
by requiring positivity of the divergence of the entropy current \cite{landau_fluid_1959}.
Such an analysis gives
\begin{eqnarray}
\nu^{\mu} & = & \sigma_{\textsc{q}}(g^{\mu\nu}+u^{\mu}u^{\nu})\left[\left(-\partial_{\nu}\mu+F_{\nu\lambda}u^{\lambda}\right)+\mu\frac{\partial_{\mu}T}{T}\right]\nonumber \\
\tau^{\mu\nu} & = & -(g^{\mu\lambda}+u^{\mu}u^{\lambda})\left[\eta(\partial_{\lambda}u^{\nu}+\partial^{\nu}u_{\lambda})+\left(\zeta-\frac{2}{D}\eta\right)\delta_{\lambda}^{\nu}\partial_{\alpha}u^{\alpha}\right].\label{eq:const}
\end{eqnarray}
The parameter $\sigma_{\textsc{q}}$ is the ``quantum critical conductivity'',
the conductivity in the absence of charges and external fields. $D$
denotes the number of spatial dimensions, which in this paper will
generically be 3.

The last term in eqs. (\ref{eq:curr}) and (\ref{eq:enTens}) are
contributions due to polarization of the material. They are present
already in thermal equilibrium. Their contribution to currents in
the bulk will be compensated by surface currents, rendering them unmeasurable
in any experimental set up \cite{cooper_thermoelectric_1997}.
Alternatively, one can derive these terms using variational techniques
such as in \cite{kovtun_thermodynamics_2016}. 

Reference \cite{hartnoll_theory_2007} presented formulas for the conductivity and resistivity tensors
within relativistic hydrodynamics in 2+1 dimensions. Here, we comment on the extension of this work to 3+1 dimensions. Unlike in $2+1$ dimensions,
where $B$ is a pseudoscalar, in higher dimensions background magnetic
fields break rotational invariance. As a consequence, we will see
that in $3+1$ dimensions, anisotropic $(\vec{E}\cdot\vec{B})B_{i}$ terms
generically arise in $J_{i}$.

\subsection{Strong Magnetic Fields}

As we mentioned in the introduction, one important caveat in the work
of \cite{hartnoll_theory_2007} and the work that followed is whether
the hydrodynamic expansion has been systematically applied. In the
constitutive relation \eqref{eq:const} we only kept terms to first
order in the gradient expansion, treating $F_{\mu\nu}$ as being first
order in the gradient expansion. The final answers hence only are
valid up to linear order in $E$ and $B$. Even assuming we may treat
$1/\tau$ as first order in derivatives, all of the novel relativistic
phenomenology of \cite{hartnoll_theory_2007} requires inclusion of
terms proportional to $\sigma_{\textsc{q}}B^{2}$ in the conductivity!
As we emphasize in this paper, there are additional hydrodynamic contributions
to the conductivity at order $B^{2}$. 

For Abelian background fields, which includes electromagnetism, one
can define an alternate expansion scheme in which the magnetic field
strength is treated as order 0 in the gradient expansion, and only
derivatives of $\vec{B}$ count as gradients. Namely, we should really
keep \textit{all} orders in $F_{\mu\nu}$ at the start of the calculation.
This is especially important for us since we are interested in NMR,
which only occurs at order $B^{2}$. Since our goal is to obtain the
linear response relation (\ref{eq:ohm}), we can consider ``weak''
electric fields and ``strong'' magnetic fields, i.e. $E\sim\mathcal{O}(\del)$,
$B\sim\mathcal{O}(1)$.    Thus, we will develop a more general constitutive
relation that keeps all orders in $\vec{B}$, but is linear in $\vec{E}$.
This is easiest to do using non-relativistic notation where $\vec{E}$
and $\vec{B}$ are treated separately.   Note that we require $B\lesssim T^2$ as a point of principle;  if this inequality is violated,  then the hydrodynamic description should be replaced by an alternative description.   For example, in the limit of extremely large $B$-fields, a hydrodynamic framework for low-lying Landau levels becomes appropriate \cite{Geracie:2014zha}.

We consider the thermodynamic quantities in (\ref{eq:curr}) and (\ref{eq:enTens})
to be functions of $\mu$, $T$ and $B^{2}$, and treat $\mu,T$ and
$u^{\mu}$ as the degrees of freedom that respond to external perturbations
to the system. The fluctuations of $\e,p$ and $\rho$ will then be
determined by the equation of state, as is standard. As in \cite{hartnoll_theory_2007},
we assume that in equilibrium the fluid velocity is $u^{\mu}=(1,\vec{0})$,
and so perturbations from equilibrium allows us to treat the velocity
of the fluid $u^{i}=v^{i}$ as $\mathcal{O}(\partial)$. Thus, any quadratic
terms in $u^{i}$ can be ignored in the gradient expansion. We also
assume that the applied electromagnetic fields are static sources;
strictly speaking, electromagnetism is a gauge theory and dynamical
gauge fields complicate the hydrodynamic description \cite{Hernandez:2017mch, Grozdanov:2016tdf}, although
the physics is only qualitatively different under extreme magnetic
fields. 

Finally, although it is not
rigorous, we will for now assume the ``mean field" approximation
that disorder modifies the momentum conservation equation simply through
the factor $1/\tau$, as in (\ref{eq:momdisp}), is correct. These
approximations hone in on the changes to the gradient expansion that
occur in a background magnetic field.  We will relax this assumption in Section \ref{sec4}.

Since without gradients and background electric field the hydrodynamic
velocity is zero, keeping only terms linear in $E$ also means keeping
only leading orders in $v$. Rotation invariance demands that the
only tensor structures allowed in $J^{i}$ are $v^{i}$, $E^{i}$,
$\epsilon_{ijk}B^{j}v^{k}$, $\epsilon_{ijk}B^{j}E^{k}$ and $B^{i}B_{j}E^{j}$.
The exact combination of these terms that is allowed to appear is
further constrained by boost invariance. Since we are working to linear
order in $E$ and hence $v$, we can restrict ourselves to Galilean
boosts which act as: \begin{subequations}
\begin{align}
t & \rightarrow t,\quad x_{i}\rightarrow x_{i}+v_{i}t,\quad J_{i}\rightarrow J_{i}+v_{i}\rho,\quad T^{ti}\rightarrow T^{ti}+(\epsilon+P)v^{i},\\
\rho & \rightarrow\rho,\quad (\epsilon+P)\rightarrow(\epsilon+P),\\
B_{i} & \rightarrow B_{i},\quad E_{i}\rightarrow E_{i}+\epsilon_{ijk}v^{j}B^{k}.
\end{align}
\end{subequations} As we are not considering inhomogeneous flows,
we can exploit boost invariance by boosting into the rest frame of
the fluid with $v^{i}=0$. In this frame the most general form of
the current reads
\begin{equation}
J^{i}=c(B^{2})E^{i}+d(B^{2})E^{j}B_{j}B^{i}+\tilde{\sigma}(B^{2})\epsilon_{\:jk}^{i}B^{j}E^{k}.
\end{equation}
The full constitutive relation from \eqref{eq:current} can be recovered
by acting on this rest frame current with a Galilean boost:
\begin{eqnarray}
\nonumber
J^{i}&=&[\rho(B^{2})+\tilde{\sigma}(B^{2})B^{2}]v^{i}-\tilde{\sigma}(B^{2})B^{i}B_{j}v^{j}+c(B^{2})(E^{i} +\e_{\:jk}^{i}v^{j}B^{k})\\ && +d(B^{2})E^{j}B_{j}B^{i}+\tilde{\sigma}(B^{2})\epsilon_{\:jk}^{i}B^{j}E^{k}.\label{eq:current}
\end{eqnarray}
The coefficients can, in general, depend on \textbf{$B^{2}$}. Finally,
we use our freedom to redefine the fluid frame to fix\footnote{In principle one could add transport coefficients analogous to $\tilde{\sigma}$,
$c$ and $d$ that appear in \eqref{eq:current} also in the momentum
current. These can however be absorbed by changing the hydrodynamic
frame, that is by redefining the velocity as $v_i\rightarrow v_i+A_1 E_i+A_2\epsilon_{ijk}B_jE_k+A_3E_jB_jB_i$.
The 3 coefficients $A_{1,2,3}$ contain enough freedom to eliminate
the 3 analogues of $\tilde{\sigma}$, $c$ and $d$ in the momentum
current.}
\begin{equation}
T^{ti}\equiv(\epsilon+P)v^{i}.\label{eq:Tti}
\end{equation}
Another way to interpret what we have found is the simple statement
that
\begin{equation}
J^{i}=\rho v^{i}+\Sigma^{ij}\left(E_{j}+\epsilon_{jkl}v^{k}B^{l}\right),\;\;\;\Sigma^{ij}\equiv c(B^{2})\delta^{ij}+\tilde{\sigma}(B^{2})\epsilon^{ikj}B_{k}+d(B^{2})B^{i}B^{j}.
\end{equation}
The first order correction to the current $J^{i}$, which was before
simply proportional to $\sigma_{\textsc{q}}$, is now proportional
to a matrix $\Sigma^{ij}$ which inherits the rotational symmetry
breaking pattern of the external magnetic field. Since entropy production,
which occurs at quadratic order in $E$, should be proportional to
$E^{i}J_{i}$, and the first term in $J^{i}$ does not contribute
to entropy production as it arises at zeroth order in hydrodynamics,
we conclude that the matrix $\Sigma^{ij}$ should be positive definite.
This constrains
\begin{equation}
c\ge0,\;\;\;c+dB^{2}\ge0.
\end{equation}
We also observe that the coefficient $\tilde{\sigma}$ is dissipationless,
and does not appear to be constrained.

The form of the constitutive relation can be further constrained if
we impose charge conjugation symmetry. Assuming that the underlying
critical theory has charge conjugation symmetry, and latter is only
broken by the explicit presence of the charge carries via $\rho$
demands symmetry under
\begin{equation}
j_{i}\rightarrow-j_{i},\quad T^{ti}\rightarrow T^{ti},\quad v^{i}\rightarrow v^{i},\quad(E^{i},B^{i})\rightarrow(-E^{i},-B^{i}),\quad\rho\rightarrow-\rho,\quad\epsilon+P\rightarrow\epsilon+P\quad
\end{equation}
This symmetry requires $\tilde{\sigma}=0$.   Generically, it is possible for $\tilde\sigma$ to be an odd function of $\rho$.   But for simplicity, we will often
set $\tilde\sigma=0$ in what follows to simplify the equations.

We confirm in Appendix
\ref{app:relativistic} that starting with the most general relativistic
constitutive relation including terms up to order $F^{3}$ indeed
yields eq. (\ref{eq:current}) up to $\mathcal{O}(B^2)$, when restricting to terms linear in
$E$. The transport coefficients $c$, $\tilde\sigma$ and $d$ appear as linear combinations
of the various terms appearing in the relativistic analysis.

\subsection{Linear Response}

We are now in a position to determine  $\sigma_{ij}$ in linear
response. The energy and charge conservation equations can be ignored
so long as the fluid is homogeneous \cite{hartnoll_theory_2007};
the momentum conservation equation reads:
\begin{equation}
(\epsilon+P)\left(-\mathrm{i}\omega+\frac{1}{\tau}\right)v^{i}\equiv\Gamma v^{i}=\rho E^{i}+\e_{\:jk}^{i}J^{j}B^{k}.\label{eq:vee}
\end{equation}
Plugging (\ref{eq:vee}) into the constitutive relation (\ref{eq:current})
gives the following matrix expression
\begin{eqnarray}
\nonumber
&& \left(\delta_{\:j}^{i}+\frac{c(B^{2})}{\Gamma}\left\{ B^{2}\delta_{\:j}^{i}-B^{i}B_{j}\right\} -\frac{\rho}{\Gamma}\e_{\:jk}^{i}B_{k}\right)J^{j}\\ &&=\left(\left\{ \frac{\rho^{2}}{\Gamma}+c(B^{2})\right\} \delta_{\:j}^{i}+\left\{ c(B^{2})\frac{\rho}{\Gamma}\right\} \e_{\,jk}^{i}B^{k}+dB^{i}B_{j}\right)E^{j}.\label{eq:matrix}
\end{eqnarray}
Without loss of generality, we let $\vec{B}=B\hat{z}$. With this,
we obtain the following expressions for the conductivity: \begin{subequations}\label{eq:sigma23}
\begin{gather}
\sigma_{zz}=\frac{\rho^{2}}{\Gamma}+c(B^{2})+dB^{2}\label{eq:conduc}\\
\sigma_{xx}=\sigma_{yy}=\Gamma\frac{\rho^{2}+\Gamma c(B^{2})+B^{2}c(B^{2})^2}{(\Gamma+c(B^{2})B^{2})^{2}+\rho^{2}B^{2}}\\
\sigma_{xy}=-\sigma_{yx}=B\rho\frac{\rho^{2}+2\Gamma c(B^{2})+B^{2}c(B^{2})^{2}}{(\Gamma+c(B^{2})B^{2})^{2}+\rho^{2}B^{2}}.
\end{gather}
\end{subequations} with the corresponding resistivity given by the
inverse matrix: \begin{subequations}  \label{eq:rho22}
\begin{gather}
\rho_{zz}=\frac{\Gamma}{d\Gamma B^{2}+\rho^{2}+\Gamma c(B^{2})}\\
\rho_{xx}=\rho_{yy}=\frac{\Gamma\text{\ensuremath{\rho}}^{2}+c(B^{2})\left(B^{2}\Gamma c(B^{2})+\Gamma^{2}\right)}{\rho^{4}+\left(\Gamma^{2}+B^{2}\rho^{2}\right)c(B^{2})^{2}+2\rho^{2}\Gamma c(B^{2})^{2}}\label{eq:resist}\\
\rho_{xy}=-\rho_{yx}=-B\frac{B^{2}\rho c(B^{2})^{2}+2\Gamma\rho c(B^{2})+\rho^{3}}{\rho^{4}+\left(\Gamma^{2}+B^{2}\rho^{2}\right)c(B^{2})^{2}+2\rho^{2}\Gamma c(B^{2})^{2}}.
\end{gather}
\end{subequations}

The constants in $c(B^{2})$ and $d$ will depend on the microscopic
details of the theory, and their sign will determine if NMR is present.
Indeed, (\ref{eq:conduc}) is reminiscent of (\ref{eq:nmrAnom}),
even though this fluid is not chiral: letting $c\approx c_{0}+c_{1}B^{2}$, and similarly for $d$,
we see that so long as $c_1+d_{0}>0$  $\sigma_{zz}$ is an increasing
function of $B^{2}$, and $\rho_{zz}$ is a decreasing function of
$B^{2}$, as is (\ref{eq:nmrAnom}). However, the main experimental
test for anomaly-induced NMR is the dramatic angular dependence of
the resistivity.   Using the definitions in (\ref{eq:resist_tensor}), it is straightforward to show that \begin{subequations}
\begin{align}
\alpha & =\frac{\Gamma\text{\ensuremath{\rho}}^{2}+c(B^{2})\left(B^{2}\Gamma c(B^{2})+\Gamma^{2}\right)}{\rho^{4}+\left(\Gamma^{2}+B^{2}\rho^{2}\right)c(B^{2})^{2}+2\rho^{2}\Gamma c(B^{2})^{2}},\\
\beta & =-B\frac{B^{2}\rho c(B^{2})^{2}+2\Gamma\rho c(B^{2})+\rho^{3}}{\rho^{4}+\left(\Gamma^{2}+B^{2}\rho^{2}\right)c(B^{2})^{2}+2\rho^{2}\Gamma c(B^{2})^{2}},\\
\gamma & =\alpha\left[-1+\dfrac{{\displaystyle 1+\dfrac{c^{2}B^{2}}{\rho^{2}+c\Gamma}\dfrac{\rho^{2}}{\rho^{2}+c\Gamma}}}{{\displaystyle \left(1+\dfrac{c^{2}B^{2}}{\rho^{2}+c\Gamma}\right)\left(1+\dfrac{d\Gamma B^{2}}{\rho^{2}+c\Gamma}\right)}}\right].
\end{align}
\end{subequations} At $B=0$, one can easily check that $\beta=\gamma=0$,
as must happen, since the theory becomes isotropic. Allowing $B\ne0$
but assuming $d=0$, we find that
\begin{equation}
\frac{\gamma}{\alpha}=-\dfrac{{\displaystyle \dfrac{c^{2}B^{2}}{\rho^{2}+c\Gamma}\dfrac{c\Gamma}{\rho^{2}+c\Gamma}}}{{\displaystyle 1+\dfrac{c^{2}B^{2}}{\rho^{2}+c\Gamma}}}<0.
\end{equation}
We hence conclude that the hydrodynamics of \cite{hartnoll_theory_2007}
can exhibit similar angular dependence in the resistivity
to (\ref{eq:nmrAnom}), despite the fact that it the hydrodynamic equations are manifestly isotropic.   This effect is dependent on the breaking of Galilean invariance, which allows for the coefficient $c\ne 0$.  Adding the $d$-dependence back in, we conclude
that only for $d$ sufficiently negative is it possible for $\gamma>0$.
At small $B$, one can show that
\begin{equation}
-d>\frac{c_{0}^{3}}{c_{0}\Gamma+\rho^{2}}  \label{eq:dNMR}
\end{equation}
is necessary in order for the angular dependence to appear as ``positive\textquotedbl{}
magnetoresistance.

If we include non-vanishing $\tilde\sigma$, then the conductivity matrix generalizes to:
\begin{subequations}\begin{align}
\sigma_{zz}&=\frac{\rho^2}{\Gamma}+c(B^{2})+d(B^{2})B^{2}\\
\sigma_{xx}&=\sigma_{yy}=\Gamma\frac{\rho\tilde{\rho}+B^{2}\tilde{\rho}\tilde{\sigma}(B^{2})+\Gamma c(B^{2})+B^{2}c(B^{2})^{2}}{(\Gamma+c(B^{2})B^{2})^{2}+\rho^{2}B^{2}}\\
\sigma_{xy}&=-\sigma_{yx}=B\frac{\tilde{\rho}^{2}\rho+\Gamma\tilde{\rho}c(B^{2})+\Gamma\rho c(B^{2})-\Gamma^{2}\tilde{\sigma}(B^{2})+B^{2}c(B^{2}){}^{2}\rho-B^{2}c(B^{2})\tilde{\sigma}(B^{2})}{(\Gamma+c(B^{2})B^{2})^{2}+\rho^{2}B^{2}}
\end{align}\end{subequations}
where $\tilde \rho = \rho - \tilde \sigma B^2$.

\subsection{Thermal Transport}

Let us now briefly discuss thermal transport. In this case, we wish
to compute the charge current and the heat current in response to
applied electric fields and thermal gradients:
\begin{equation}
\left(\begin{array}{c}
J^{i}\\
Q^{i}
\end{array}\right)=\left(\begin{array}{cc}
\sigma^{ij} & \ T\alpha^{ij}\\
T\bar{\alpha}^{ij} & \ T\bar{\kappa}^{ij}
\end{array}\right)\left(\begin{array}{c}
E_{j}\\
-T^{-1}\partial_{j}T
\end{array}\right)
\end{equation}
The heat current is defined as \cite{hartnoll_theory_2007}
\begin{equation}
Q^{i}\equiv T^{ti}-\mu J^{i}  \label{eq:heatcurrent}
\end{equation}
with $\mu$ the chemical potential of the fluid.

A priori, such a computation can be quite subtle, since it appears
as though we need to account for more terms in the derivative expansion
to fix the constitutive relations for $\partial_{j}T$. However, consider
the following arguments. Firstly, we may use the standard Landau frame
in which (\ref{eq:Tti}) is exact. Using the thermodynamic relation
$\epsilon+P-\mu\rho=Ts$, with $s$ the entropy density, we conclude
that in an electric field $E^{i}$ (but keeping $\partial_{i}T=0$):
\begin{equation}
Q^{i}=Tsv^{i}-\mu\Sigma^{ij}(E_{j}+\epsilon_{jkl}v^{k}B^{l}).
\end{equation}
We have already solved for the velocity field $v^{i}$ in an applied
electric field in our computation of $\sigma^{ij}$, so hence we obtain
straightforwardly the matrix \\
\begin{gather}
\bar{\alpha}_{xx}=\frac{\Gamma\rho sT-c\Gamma\mu\left(B^{2}c+\Gamma\right)}{T\left(\left(B^{2}c+\Gamma\right)^{2}+B^{2}\rho^{2}\right)}\label{eq:baralpha-1}\\
\bar{\alpha}_{xy}=\frac{B\left(csT\left(B^{2}c+\Gamma\right)-c\Gamma\mu\rho+\rho^{2}sT\right)}{T\left(\left(B^{2}c+\Gamma\right)^{2}+B^{2}\rho^{2}\right)}\nonumber \\
\bar{\alpha}_{zz}=\frac{\rho s}{\Gamma}-\frac{\mu\left(B^{2}d+c\right)}{T}.\nonumber
\end{gather}

Secondly, we use Onsager reciprocity which states that $\alpha_{ij}(B)=\bar{\alpha}_{ji}(-B)$;
since all off-diagonal elements of the transport matrices are antisymmetric,
we conclude that $\alpha_{ij}=\bar{\alpha}_{ij}$. Next, we can imagine
turning off the eletric field and only applying an external temperature
gradient. The momentum conservation equation then reads
\begin{equation}
\Gamma v^{i}=\epsilon^{ijk}J_{j}B_{k}-Ts\frac{\partial^{i}T}{T}.  \label{eq:24mom}
\end{equation}
Using the fact that
\begin{equation}
J^{i}=-\bar{\alpha}_{ij}\del^{j}T,  \label{eq:24J}
\end{equation}
we may combine (\ref{eq:heatcurrent}), (\ref{eq:24mom}) and (\ref{eq:24J}) to obtain $\bar{\kappa}_{ij}$: \begin{subequations}
\begin{gather}
\bar{\kappa}_{xx}=\frac{B^{2}c\left(c\Gamma\mu^{2}+(\mu\rho+sT)^{2}\right)+\Gamma\left(c\Gamma\mu^{2}+s^{2}T^{2}\right)}{T\left(\left(B^{2}c+\Gamma\right)^{2}+B^{2}\rho^{2}\right)}\\
\bar{\kappa}_{xy}=\frac{BsT(\rho sT-2c\Gamma\mu)-B^{3}c^{2}\mu(\mu\rho+2sT)}{T\left(\left(B^{2}c+\Gamma\right)^{2}+B^{2}\rho^{2}\right)}\\
\bar{\kappa}_{zz}=\frac{\mu^{2}\left(B^{2}d+c\right)}{T}+\frac{s^{2}T}{\Gamma}.
\end{gather}
\end{subequations} 
From these results we can conclude that the constitutive relation for the current must include a linear temperature gradient as \begin{equation}
J^i = \rho v^i + \Sigma^{ij}\left(E^j + \epsilon^{jkl}v_k B_l - \frac{\mu}{T}\partial^jT\right).
\end{equation}
As in an ordinary relativistic fluid, we conclude that there are no new dissipative coefficients associated with thermo-magnetic response.

\section{``Microscopic''  Examples}

\label{sec3}

\subsection{$\protect\susn=4$ SYM Plasma}

We now wish to compare our formalism with the conductivity of $N_{\mathrm{f}}$
massive $\susn=2$ supersymmetric hypermultiplets flowing through an $\susn=4$
SYM plasma with gauge group $\mathrm{SU}(N_{\mathrm{c}})$ at temperature $T$. This
model was studied extensively starting with \cite{Karch:2007pd} and the conductivities in the background of constant electromagnetic field with generic orientations was worked out in \cite{ammon_holographic_2009}. We take the limits $N_{\mathrm{c}}\to\infty$ with large but
finite 't Hooft coupling $\lambda=g_{\mathrm{YM}}^{2}N_{\mathrm{c}}$, allowing the use
of holographic techniques. The flavor hypermultiplets are dual to $N_{\mathrm{f}}$
D7 branes \cite{Karch:2002sh} embedded in a fixed AdS-Schwarzschild
background. Furthermore, we will work in the probe limit $N_{\mathrm{f}}\ll N_{\mathrm{c}}$
so that we may neglect the back reaction of the probe branes on the
supergravity fields. This allows us to treat the plasma as stationary,
and focus on the dynamics of the flavor fields alone. Specifically,
this limit allows for an apparent dissipation of momentum. The flavor
fields lose energy to the plasma at a rate of order $N_{\mathrm{c}}$ so only
at times of order $N_{\mathrm{c}}$ will the back reaction on the $N_{\mathrm{c}}^2$ plasma degrees of freedom be
non-negligible.

This momentum relaxation is rather `peculiar', and so the theory of transport in probe brane models differs in important ways from other models of transport \cite{rmp}.  In particular, it is unclear whether or not a `weak disorder' limit exists.    As such a limit was required in order to rigorously include $\Gamma$ in our hydrodynamic model of transport, there is  a priori no reason to expect exact quantitative agreement between our hydrodynamic model and this holographic model.    It is known that generic holographic models disagree with the hydrodynamics of \cite{Davison:2015bea, Blake:2015epa, Blake:2015hxa} at next-to-leading order in $\Gamma$:   this can crudely be thought of as arising due to $\Gamma$-dependent corrections to the hydrodynamic constitutive relations.

The conductivity of the propagating hypermultiplets in generic constant background fields was found in \cite{ammon_holographic_2009}.
They considered an $E$ field that is fixed along the $x$-axis, while
the $B$ field lies in the $x-z$ plane. This can be mapped onto our
formalism by rewriting their results in terms of the basic constitutive
relation \eqref{eq:resist_tensor}. For small electric fields, they
found
\begin{gather}
\sigma_{xx}=\tilde{\rho}\frac{1+b_{x}^{2}}{1+b^{2}}\sqrt{1+\frac{N_{\mathrm{f}}^{2}N_{\mathrm{c}}^{2}T^{2}}{\tilde{\rho}^{2}16\pi^{2}}\cos^{6}\theta^{\star}(1+b^{2})},\quad\sigma_{xy}=\frac{\tilde{\rho}b_{z}}{1+b^{2}},\quad\sigma_{xz}=\frac{b_{x}b_{z}}{1+b_{x}^{2}}\sigma_{xz}\label{eq:d7cond}
\end{gather}
where $b_{i}=\frac{B_{i}}{qT^{2}}$ and $\tilde{\rho}=\frac{\rho}{qT^{2}}$
. This can be brought into the form \eqref{eq:resist_tensor} with
coefficients
\begin{subequations}\label{eq:d7coeff}\begin{align}
\alpha&=\frac{q\rho T^{2}\sqrt{\frac{2\sigma_{0}T^{2}\left(B^{2}+q^{2}T^{4}\right)}{q\rho^{2}}+1}}{\rho^{2}+2q\sigma_{0}T^{6}} \\
\beta &=-\frac{\rho}{\rho^{2}+2q\sigma_{0}T^{6}} \\
\gamma &=\frac{qT^{2}}{\rho\sqrt{\frac{2\sigma_{0}T^{2}\left(B^{2}+q^{2}T^{4}\right)}{q\rho^{2}}+1}}-\alpha
\end{align}\end{subequations}
where $\sigma_{0}=\frac{N_{\mathrm{f}}^{2}N_{\mathrm{c}}^{2}\sqrt{\lambda}}{64\pi}\cos^{6}\theta^{\star}$
and $q=\frac{\pi}{2}\sqrt{\lambda}$.

This results look quite different from the hydrodynamic form. One could ask whether the D3/D7 answer can be fit into the hydrodynamic framework by a particular choice of transport coefficients.
At large $\rho$, the coefficients $c$, $d$ and $\tilde{\sigma}$ are generically $\rho$-dependent.    In the limits $\rho \gg T^3$ and $B\ll qT^2$  limits,   \cite{ammon_holographic_2009} have
shown that probe brane models appear consistent with
\[
\Gamma=qT^{2}\rho
\]
at leading order in $\rho$.   At this order, one trivially finds a Drude-like conductivity: $\sigma_{ij} \approx \rho^2 \delta_{ij}/\Gamma$.    Since $c$, $d$ and $\tilde{\sigma}$ arise at next-to-leading order in this limit, their unique determination requires specifying $\Gamma$ at order $\rho^0$.  It is unclear whether this question is even `well-posed', in light of the subtleties that arise in transport beyond the weak disorder limit \cite{Davison:2015bea, Blake:2015epa, Blake:2015hxa, rmp}.


However, given the exact magnetic field dependence of the resistivity, we can non-perturbatively compute the magnetoresistance in both $B$ and $\rho$.   Firstly, one can explicitly compute \begin{equation}
\rho_{zz} = \frac{qT^2}{\sqrt{\rho^2 +\frac{2}{q}T^2\sigma_0 \left(B^2+q^2T^4\right)}},
\end{equation}
which is clearly a decreasing function of $B^2$.
 Secondly, the ratio of the longitutinal to transverse
resistivities is
\begin{equation}
\frac{\rho_{xx}}{\rho_{zz}}=1+\frac{2\sigma_{0}T^{2}}{q(\rho^{2}+2q\sigma_{0}T^{6})}B^{2}>1\label{eq:d7NMR}
\end{equation}
implying the resistivity along the direction of the magnetic field
is supressed relative to the transverse directions.   As expected, in the $B\to0$
limit, the ratio goes to one since the theory becomes isotropic. \\
\\

\subsection{Cartoon of electron-hole plasma}

In this section we present a simple classical cartoon of a fluid where
we can compute the coefficients $c$ and $d\ne0$.   More precisely,  let
us consider a toy model of two charged fluids, one of charge density
$\hat{\rho}$ (the $+$ fluid) and the other of charge density $-\hat{\rho}$
(the $-$ fluid), analogous to electron and hole fluids in graphene.  Unlike in
graphene \cite{2009PhRvB..79h5415F}, we will suppose that the momentum of these two charged
fluids is also an almost conserved quantity. For simplicity, we assume
that all other properties, such as enthalpy $\hat{\mathcal{M}}$,
of these two fluids are identical, and we also only consider the hydrodynamic
gradient expansion to first order in derivatives.

The net current is given by the sum of currents in the $\pm$ fluids:
$J^{\mu}=J_{+}^{\mu}+J_{-}^{\mu}$. The spatial components of these
currents are given by \begin{subequations}\label{eq:2FJ}
\begin{align}
J_{+}^{i} & =\hat{\rho}v_{+}^{i}+\hat{\sigma}_{\textsc{q}}\left(E^{i}+\varepsilon^{ijk}v_{+}^{j}B^{k}\right),\\
J_{-}^{i} & =-\hat{\rho}v_{-}^{i}+\hat{\sigma}_{\textsc{q}}\left(E^{i}+\varepsilon^{ijk}v_{-}^{j}B^{k}\right).
\end{align}
\end{subequations} $\hat{\sigma}_{\textsc{q}}$ is the `quantum critical
conductivity' for each microscopic fluid, and will be important to
include. The momentum quasi-conservation equations of the two fluids
are \begin{subequations}\label{eq:2FP}
\begin{align}
-\mathrm{i}\omega\hat{\mathcal{M}}v_{+}^{i} & =\hat{\rho}E^{i}+\varepsilon^{ijk}J_{+}^{j}B^{k}-\alpha(v_{+}^{i}-v_{-}^{i}),\\
-\mathrm{i}\omega\hat{\mathcal{M}}v_{-}^{i} & =-\hat{\rho}E^{i}+\varepsilon^{ijk}J_{-}^{j}B^{k}-\alpha(v_{-}^{i}-v_{+}^{i}).
\end{align}
\end{subequations} $\alpha/\hat{\mathcal{M}}$ governs the rate at
which the electron/hole fluids exchange momentum. (\ref{eq:2FJ})
and (\ref{eq:2FP}) form a set of equations which can be solved straightforwardly:
upon doing so, we find that \begin{subequations}\label{eq:2FC}
\begin{align}
\nonumber
\sigma_{xx}(\omega) & =\frac{-2\mathrm{i}\omega\hat{\mathcal{M}}
\left(\hat{\rho}^{2}+\hat{\sigma}_{\textsc{q}}\left(2\alpha+B^{2}\hat{\sigma}_{\textsc{q}}-\mathrm{i}
\omega\hat{\mathcal{M}}\right)\right)}{B^{4}\hat{\sigma}_{\textsc{q}}^{2}-2\mathrm{i}\alpha\hat{\mathcal{M}}
\omega-(\omega\hat{\mathcal{M}})^{2}+B^{2}\left(\hat{\rho}^{2}
+2\hat{\sigma}_{\textsc{q}}\left(\alpha-\mathrm{i}\omega\hat{\mathcal{M}}\right)\right)} \\
&\approx\frac{-2\mathrm{i}\omega\hat{\mathcal{M}}\left(\hat{\rho}^{2}+\hat{\sigma}_{\textsc{q}}\left(2\alpha+B^{2}\hat{\sigma}_{\textsc{q}}\right)\right)}{B^{4}\hat{\sigma}_{\textsc{q}}^{2}-2\mathrm{i}\alpha\hat{\mathcal{M}}\omega+B^{2}\left(\hat{\rho}^{2}+2\hat{\sigma}_{\textsc{q}}\alpha\right)},\\
\sigma_{zz}(\omega) & =2\frac{\hat{\rho}^{2}+(2\alpha-\mathrm{i}\hat{\mathcal{M}}\omega)\hat{\sigma}_{\textsc{q}}}{2\alpha-\mathrm{i}\omega\hat{\mathcal{M}}}\approx2\hat{\sigma}_{\textsc{q}}+\frac{\hat{\rho}^{2}}{\alpha}.
\end{align}
\end{subequations} The parameter $\alpha$ governs the rate of relaxation
between the two fluids. Assuming that $\alpha$ is small enough that
it can be treated within the gradient expansion of hydrodynamics,
one can show using that the second law of thermodynamics implies
\begin{equation}
\alpha>0.\label{eq:2FA}
\end{equation}
Perhaps more intuitively, (\ref{eq:2FA}) can also be understood from
the requirement that thermal equilibrium $v_{\pm}^{i}=0$ is stable.

The approximations we have made in the last step of (\ref{eq:2FC})
are valid in the limit $\alpha\gg\omega\hat{\mathcal{M}}$. In this
limit, we expect single fluid hydrodynamics with net charge density
zero. From (\ref{eq:sigma23}) (with $\rho=0$ and $\Gamma=-\mathrm{i}\omega\mathcal{M}$)
we should find \begin{subequations}
\begin{align}
\sigma_{xx}(\omega) & =\frac{-\mathrm{i}\omega\mathcal{M}\sigma_{\perp}}{B^{2}\sigma_{\perp}-\mathrm{i}\omega\mathcal{M}}+\mathcal{O}\left(\omega^{3}\right),\\
\sigma_{zz}(\omega) & =\sigma_{\parallel}+\mathcal{O}(\omega)
\end{align}
\end{subequations} with $\sigma_{\perp}=c$ and $\sigma_{\parallel}=c+dB^{2}$.
Upon making the identifications \begin{subequations}
\begin{align}
\mathcal{M} & =2\hat{\mathcal{M}},\\
\sigma_{\parallel} & =2\hat{\sigma}_{\textsc{q}}+\frac{\hat{\rho}^{2}}{\alpha}\\
\sigma_{\perp} & =\sigma_{\parallel}+\frac{B^{2}\hat{\sigma}_{\textsc{q}}}{\alpha},
\end{align}
\end{subequations} we see that (\ref{eq:sigma23}) and (\ref{eq:2FC}) agree. Furthermore,
we find an expression for
\begin{equation}
d=-\frac{\hat{\sigma}_{\textsc{q}}}{\alpha}.
\end{equation}
From (\ref{eq:2FA}), we conclude that $d<0$.    Using (\ref{eq:dNMR}), we see that this model will exhibit positive magnetoresistance if momentum relaxation is strong enough.

\section{Momentum Relaxation Rate}

\label{sec4}

So far, we have used a `mean field' description of momentum relaxation.   It is also possible that upon adding a magnetic field,  momentum can relax more or less efficiently in the direction of the magnetic field.   In this section, we will perturbatively compute the rate of momentum relaxation in a fluid, disordered by very long wavelength inhomogeneity in an externally imposed chemical potential \cite{Lucas:2015lna, Lucas:2015sya}.    In such a limit, the transport coefficients may be computed by solving the hydrodynamic equations in an inhomogeneous medium, which can be shown to be:  \begin{subequations}\label{eq:inhomohydro}\begin{align}
\partial_i J^i = \partial_i \left(\rho(\mu(x)) \delta v^i - \Sigma^{ij}\left(\partial_j \delta \mu - \frac{\mu(x)}{T}\partial_j \delta T(x) - \epsilon_{jkl}v^k B^l\right)\right) &= 0, \\
\partial_i \left(Ts(\mu(x)) \delta v^i - \mu(x) \Sigma^{ij}\left(\partial_j \delta \mu - \frac{\mu(x)}{T}\partial_j \delta T(x) - \epsilon_{jkl}v^k B^l\right)\right) &= 0, \\
\rho(\mu(x))\partial_i \delta \mu + s(\mu(x))\partial_i \delta T - \partial_j \left(\eta \left(\partial_i \delta v_j + \partial_j \delta v_i - \frac{2}{3}\delta_{ij}\partial_k\delta v_k\right)\right) &= \epsilon_{ijk}J^jB^k.
\end{align}\end{subequations}
Suppose that $\mu(x) = \bar\mu + u \hat\mu(x)$, with $u$ perturbatively small.   One can compute $\Gamma$ to leading order in $u$ by either solving (\ref{eq:inhomohydro}) in an inhomogeneous background \cite{Lucas:2015lna, Lucas:2015sya, KS11}, or by using the memory function formalism \cite{forster1995, Hartnoll:2012rj, Lucas:2015pxa, rmp}.   For our purposes, it will be easier to do the latter.   What one finds is that $\Gamma v_i$ in (\ref{eq:vee}) should be replaced by $\Gamma_{ij}v_j$, with \begin{equation}
\Gamma_{ij} \equiv \int \frac{\mathrm{d}^3k }{(2\pi)^3} k_i k_j |\mu(k)|^2 \times \lim_{\omega\rightarrow0} \frac{\mathrm{Im}\left(G^{\mathrm{R}}_{\rho\rho}(\omega,k)\right)}{\omega},  \label{eq:Gij}
\end{equation}
with the retarded Green's function evaluated in the translation invariant theory, and $\mu(k)$ the Fourier transform of $\mu(x)$.     The only non-zero contributions to $\Gamma_{ij}$ will be at least $\mathcal{O}(u^2)$.   For simplicity in what follows, we will neglect the anisotropic corrections to the `quantum critical' conductivity $\sigma_{\textsc{q}}$ which can arise in a magnetic field.

The hydrodynamic Green's functions may be found by the following prescription \cite{Kovtun:2012rj}.   Suppose the hydrodynamic equations of motion take the schematic form \begin{equation}
\partial_t \varphi_A + M_{AB}\varphi_B = 0.
\end{equation}
Let $\chi_{AB}$ be the susceptibility matrix:  $\chi_{AB} = \partial \varphi_A/\partial \lambda_B$, with $\lambda_B$ the thermodynamic conjugate variables to $\varphi_A$.    For us,  $\varphi_A = (\epsilon, \rho, T^{ti})$ and $\delta \lambda_A = (\delta T/T,  \delta \mu - \mu \delta T /T,  \delta v_i)$,  and \cite{Kovtun:2012rj}\begin{equation}
\chi_{AB} = \left( \begin{array}{ccc} T(\partial_T\epsilon)_{\mu/T} &\ (\partial_\mu \epsilon)_T &\ 0 \\ (\partial_\mu\epsilon)_T &\ (\partial_\mu n)_T &\ 0 \\ 0 &\ 0 &\ (\epsilon+P)\delta_{ij}  \end{array}\right).
\end{equation}
The hydrodynamic retarded Green's function is \begin{equation}
G^{\mathrm{R}}_{AB}(k,\omega) = \left[M(k) ((M(k)-\mathrm{i}\omega)^{-1} \chi \right]_{AB}.
\end{equation}
From the equations of motion in a magnetic field, we find (neglecting viscous effects, for simplicity, as these are subleading in the limit where $\hat\mu(x)$ is extremely slowly varying \cite{Lucas:2015lna, Lucas:2015sya, KS11}) \begin{equation}
M_{AB} = \left(\begin{array}{ccc} 0 &\ 0 &\ \mathrm{i}k_i \\ \sigma_1 k^2 &\ \sigma_2 k^2 &\ \mathrm{i}k_j \frac{\rho \delta_{ij} + \sigma_{\textsc{q}}\epsilon_{ijk}B_k}{\epsilon+P} \\ \mathrm{i}k_i b_1 + \mathrm{i}\sigma_1 \epsilon_{ijk}k_j B_k &\   \mathrm{i}k_i b_2 + \mathrm{i}\sigma_2 \epsilon_{ijk}k_j B_k &\  \epsilon_{imk}B_k \left(\frac{\rho}{\epsilon+P} \delta_{jm} + \frac{\sigma_{\textsc{q}}}{\epsilon+P}\epsilon^{mjn}B_n\right)  \end{array}\right),
\end{equation}with the parameters \begin{subequations}\begin{align}
\frac{\sigma_1}{\sigma_{\textsc{q}}} &= \left(\frac{\partial \mu}{\partial\epsilon}\right)_\rho - \frac{\mu}{T}\left(\frac{\partial T}{\partial \epsilon}\right)_\rho, \;\;\;\; \frac{\sigma_2}{\sigma_{\textsc{q}}} = \left(\frac{\partial \mu}{\partial\rho}\right)_\epsilon - \frac{\mu}{T}\left(\frac{\partial T}{\partial \rho}\right)_\epsilon, \\
b_1 &= \left(\frac{\partial P}{\partial \epsilon}\right)_\rho, \;\;\;\;\; b_2 = \left(\frac{\partial P}{\partial \rho}\right)_\epsilon.
\end{align}\end{subequations}


Using these equations, we now compute the $(k,\omega)\rightarrow 0$ limit of $G^{\mathrm{R}}(k,\omega)$.   What we find is that the spectral weight then diverges as $k^{-2}$: \begin{equation}
\lim_{\omega\rightarrow} \frac{\mathrm{Im}\left(G^{\mathrm{R}}_{\rho\rho}(\omega,k)\right)}{\omega} = \mathcal{A} \frac{\rho^2 + c^2B^2 }{c \left[k^2\left(\rho^2 + 2c^2B^2\right) - c^2 (k_iB_i)^2\right] }  + \cdots,
\end{equation}
where the thermodynamic prefactor \begin{equation}
\mathcal{A} = \frac{(\partial_\epsilon P)_\rho (\rho (\partial_\mu \epsilon)_T - (\epsilon+P)(\partial_\mu \rho)_T)}{T(\epsilon+P)((\partial_\rho P)_\epsilon (\partial_\epsilon \frac{\mu}{T})_\rho - (\partial_\epsilon P)_\rho (\partial_\rho \frac{\mu}{T})_\epsilon)}.
\end{equation}
It is straightforward to see that spectral weight is enhanced by a magnetic field parallel to the wave vector.   Hence, assuming $\mu(k)$ is random, upon performing the angular integral in (\ref{eq:Gij}), we find that $\Gamma_{zz} > \Gamma_{xx}=\Gamma_{yy}$.     Since $\Gamma_{zz}/\Gamma_{xx} >1$,  this type of inhomogeneity tends towards `positive magnetoresistance' by causing $\rho_{zz}/\rho_{xx}$ to increase, relative to the scenario with $\Gamma$ homogeneous.

\section{Conclusion}

\label{sec5}

In this article we have presented a relativistic hydrodynamic theory of magnetotransport in $3+1$ dimensions.    Depending on the `microscopic' models of interest, it is possible to obtain both positive and negative magnetoresistance within our framework.  NMR is, in some ways, a more generic effect for a relativistic or non-Galilean invariant fluid:  arising not only from the spatial anisotropies caused by the presence of background magnetic fields, but also from the particular structure of relativistic hydrodynamics.        In particular, we found that the holographic D3/D7 system exhibited negative magnetoresistance.   A two-fluid cartoon of electron-hole plasma which exhibits positive magnetoresistance can also be found.    Smooth disorder potentials also imply a slight positive magnetoresistance.

In Galilean invariant fluids, it has been shown \cite{alekseev2016, andreev2016, 2017arXiv170307325S} that magnetoresistance is sensitive only to viscosity.   However, we have shown in Section \ref{sec4} that for other gradient expansions, this no longer remains the case.   It is not clear whether  magnetoresistance is a good viscometer for electron fluids arising from general band structures.   This may not be the case, unless the temperature is low enough that thermal effects are negligible.   Further work to resolve this question is warranted.


\section*{Acknowledgements}
We thank Anton Andreev for comments on a draft of this paper.
The work of AK is supported, in part, by the US Department of Energy
under grant number DE-SC0011637. AL is supported by the Gordon and
Betty Moore Foundation.

\appendix

\section{Constitutive Relations to Third Order in $F_{\mu\nu}$}

\label{app:relativistic}

Here we show that the non-relativistic constitutive relations can
be derived from the most general covariant expression up to third
order in the field strength. The expression is
\begin{align}
J^{\mu}&=\rho u^{\mu}+\sigma_{\textsc{q}1}F_{\lambda}^{\mu}u^{\lambda}+\sigma_{\textsc{q}2}F^{\mu\rho}F_{\rho\sigma}u^{\sigma} +\sigma_{\textsc{q}3}F^{2}F_{\:\rho}^{\mu}u^{\rho}+\sigma_{\textsc{q}4}F^{\mu\rho}F_{\rho\sigma}F_{\:\lambda}^{\sigma}u^{\lambda} \notag \\
&+\sigma_{\textsc{q}5}F_{\rho}^{\mu}F_{\:\lambda}^{\sigma}F_{\sigma\eta}u^{\rho}u^{\lambda}u^{\eta}+\sigma_{\textsc{q}6}F_{\:\rho}^{\mu}F_{\:\sigma}^{\rho}\left(\star F\right)_{\:\nu}^{\sigma}u^{\nu}
+\sigma_{\textsc{q}7}\e^{\mu\nu\rho\sigma}F_{\nu\tilde{\nu}}F_{\rho\sigma}u^{\tilde{\nu}} \notag \\
&+\sigma_{\textsc{q}8}F_{\:\rho}^{\mu}\left(\star F\right)_{\:\sigma}^{\rho}\left(\star F\right)_{\:\nu}^{\sigma}u^{\nu}+\sigma_{\textsc{q}9}(\star F)_{\rho\sigma}F^{\rho\sigma}F_{\:\nu}^{\mu}u^{\nu}+\sigma_{\textsc{q}10}(\star F)_{\rho\sigma}F^{\rho\sigma}(\star F)_{\:\nu}^{\mu}u^{\nu}.
\end{align}
We will only keep track of terms to ``first order" in $E$, recalling that $v_i$ is first order in $E$ while $B$ is zeroth order. The first term simply gives
\begin{equation}
\sigma_{\textsc{q}1}F_{\lambda}^{\mu}u^{\lambda}\to\sigma_{\textsc{q}1}(E^{i}+\e^{ijk}v_{j}B_{k})
\end{equation}
which is expected by Galilean invariance. The second term gives
\begin{align}
\sigma_{\textsc{q}2}F^{\mu\rho}F_{\rho\sigma}u^{\sigma}&\rightarrow\sigma_{\textsc{q}2}\left(\e^{ijn}\e_{jkm}B_{n}B_{m}v^{k}-\e^{ijk}E_{j}B_{k}\right) \notag \\
&=\sigma_{\textsc{q}2}\left\{ \left(\d_{\:k}^{n}\d_{\:m}^{i}-\d_{\:m}^{n}\d_{\:k}^{i}\right)B_{n}B_{m}v^{k}-\e^{ijk}E_{j}B_{k}\right\} \notag \\
&=\sigma_{\textsc{q}2}\left((v\cdot B)B^{i}-B^{2}v^{i}-\e^{ijk}E_{j}B_{k}\right).
\end{align}
The third term gives a term identical to the first but with a $B^{2}$:
\begin{equation}
\sigma_{\textsc{q}3}F^{2}F_{\:\rho}^{\mu}u^{\rho}\rightarrow2\sigma_{\textsc{q}3}B^{2}\left(E^{i}+\e^{ijk}v_{j}B_{k}\right).
\end{equation}
The fourth term gives \begin{equation}
\sigma_{\textsc{q}4}F^{\mu\rho}F_{\rho\sigma}F_{\:\lambda}^{\sigma}u^{\lambda} \rightarrow -\sigma_{\textsc{q}4} \left(B^2\delta_{ij}-B_i B_j\right) \left(E^{i}+\e^{ijk}v_{j}B_{k}\right).
\end{equation}
The fifth, sixth and seventh terms contribute at $\mathcal{O}(E^{2}).$
The eighth term gives
\begin{equation}
\sigma_{\textsc{q}8}F_{\:\rho}^{\mu}\left(\star F\right)_{\:\sigma}^{\rho}\left(\star F\right)_{\:\nu}^{\sigma}u^{\nu}\to\sigma_{\textsc{q}8}(-B^{2}E^{i}+B^{2}E_{i}-(E\cdot B)B^{i}).
\end{equation}
The term $(\star F)_{\rho\sigma}F^{\rho\sigma}$ in the ninth and
tenth terms gives $-4E\cdot B$, and so \begin{subequations}\begin{align}
\sigma_{\textsc{q}9}(\star F)_{\rho\sigma}F^{\rho\sigma}F_{\:\nu}^{\mu}u^{\nu}&\rightarrow-4\sigma_{\textsc{q}9}(E\cdot B)(E^{i}+\e^{ijk}v_{j}B_{k}) \\
\sigma_{\textsc{q}10}(\star F)_{\rho\sigma}F^{\rho\sigma}(\star F)_{\:\nu}^{\mu}u^{\nu}&\rightarrow4\sigma_{\textsc{q}10}(E\cdot B)B^{i}.
\end{align}\end{subequations}
Only the latter term contributes at linear order in $E$ and $v$.  Collecting these equations, we find that \begin{subequations}\begin{align}
c &= \sigma_{\textsc{q}1} + \left(2\sigma_{\textsc{q}3} - \sigma_{\textsc{q}4}\right)B^2 + \cdots, \\
d &= \sigma_{\textsc{q}4}+4\sigma_{\textsc{q}10}-\sigma_{\textsc{q}8} + \cdots.
\end{align}\end{subequations}

%
%
%
%
%

\bibliographystyle{JHEP}
\bibliography{NMR}

\end{document}